\documentclass[]{zakoproc}
\usepackage{graphicx}
\begin{document}

\title{Modelling of the free-free absorption in the starburst galaxy M82 and the Sgr A Complex}
\author{W. S. Jurusik, K. T. Chy\.zy, R. T. Drzazga}
\institute{Astronomical Observatory of the Jagiellonian University, ul. Orla 171, 30-244 Krak\'ow,
Poland}
\markboth{W. S. Jurusik et al.}{Modelling of the free-free absorption in the starburst galaxy M82 and the Sgr A Complex \ldots}

\maketitle

\begin{abstract}
The radio emission of normal galaxies may become opaque at low radio frequencies
due to thermal ionized gas. We performed modelling of the free-free absorption to 
reproduce the local spectrum of SgrA Complex and the global spectrum of the starburst galaxy M82. 
We show the importance of resolution of radio observations and the value of filling factor of the 
absorbing gas for correct modelling of the absorption.
\end{abstract}

\section{Local spectrum of Sgr A}
 The radio emission of normal galaxies is directly connected with activity of star formation and results from two processes, non-thermal (synchrotron) and thermal (free-free) radiation.
 The synchrotron emission dominates the $1-10$\,GHz range. At higher frequencies free-free emission becomes increasingly important. At much lower radio frequencies free-free absorption may affect not only thermally generated photons but also synchrotron ones. Due to this absorption the radio spectra of galaxies can be flattened towards low frequencies.

The Galactic Centre (GC) hosts  a compact non-thermal radio source known as Sgr A*. 
Around Sgr A*  radio emission comes from  HII region called Sgr A West, embedded in lower density ionized gas forming extended ($1.5\arcmin$) halo (Pedlar el al. 1989).
Behind the Sgr A West there is a source, which seems to be a supernova remnant (SNR) known as Sgr A East. Around Sgr A East there is a $7\arcmin$ radio envelope, showing thermal and non-thermal emission.
At low radio frequencies ($<$610MHz) the most striking feature of the radio emission of Sgr A  is  its strong depression in the region of Sgr A West which at higher frequencies dominates the radio image. 

In modelling of the radio spectrum of the whole Sgr A Complex consisting of Sgr A West, Sgr A East, and Sgr A*, we assume three gaseous components of different sizes and radiating according to thermal and/or  synchrotron processes.
The emission of all components is modelled over a grid of NxN cells. Each simple cell 
radiates according to:
$$S_{ij}=[S_{sync,ij}\,\nu^{-\alpha}\,exp(-\tau_{\nu,ij})+S_{th,ij}\,(1-\exp(-\tau_{\nu,ij}))]\, \Omega_{ij}$$

where $\tau_{\nu,ij}=0.0824(\frac{T}{K})^{-1.35}(\frac{\nu}{GHz})^{-2.1}EM$ is the optical depth  and $EM=\int_0^l N_e^2\,f_{\nu}\,dl$ is the emission measure with the filling factor $f_{\nu}$.

The modelled spectrum of Sgr A integrated over the whole complex fits very well the data points (Fig. 1: 
Left).The modelled turnover in the spectrum of Sgr A near 1 GHz requires 
emission measures $> 10^7\, pc\, cm^{-6}$. This provides the thermal 
emission contaminating the high frequency part of the spectrum. To explain 
the data points it was also necessary to introduce the filling factor $f_\nu$ 
of about 0.1 in the part of the model associated with the thermal emission.

\begin{figure}[ht]
 \includegraphics[width=.32\textwidth]{spectrum.eps}
  \includegraphics[width=.32\textwidth]{m82spectrum.eps}
   \includegraphics[width=.35\textwidth]{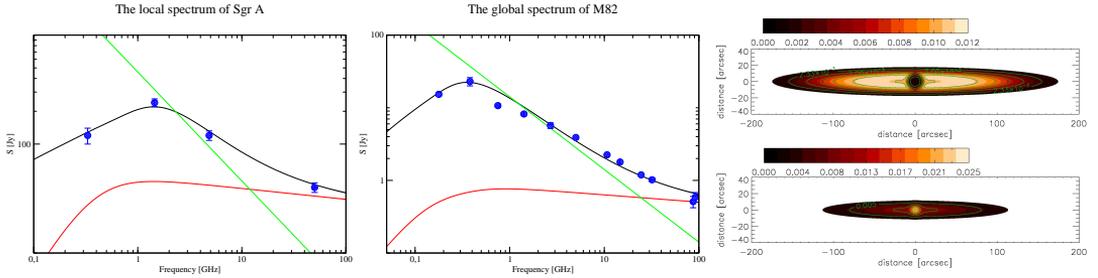}
 \small {\caption{ ({\bf Left}) The modelled spectrum of Sgr A Complex. 
 The three lines correspond to total, synchrotron, and thermal spectrum. The data are from  Pedlar et al. 1989. ({\bf Centre}) The global spectrum of M82. The data are from  Klein et al. 1988. ({\bf Right}) The modelled map of radio emission of M82 at 300MHz (top), and 1400 MHz (bottom).}}
 \label{description}
\end{figure}

\section{Global spectrum of M82}

The galaxy M82 hosts the nearest example of an ongoing starburst. The brightest source at 408\,MHz (Wills et al. 1997) is a SNR in the centre of the disk. Around this SNR there is a depression in radio emission of about 100\,pc in size caused by free-free absorption induced by a giant HII region.

We performed modelling of the global spectrum of M82 similarly to Sgr A complex, but for the whole galaxy (Figure 1:Centre). The shape of low frequency spectrum depends on the distribution and density of the ionized gas in the whole galactic disc.  At fixed frequencies, we  reconstructed the observed radio emission over the galactic disk. Figure 1: Right depicts some modelled maps at 300 and 1400 MHz.
With lower resolution the depression in emission in the centre of M82 is systematically less visible, at 408 MHz and with spatial resolution of $10\arcsec$ it even disappears completely.  Hence, recognition of such regions in more distant galaxies would be very difficult.

Our modelling shows that significant part of M82 (more than 50\%) must 
be significantly contaminated by thermally emitting gas. It is not possible 
to reproduce the turnover in the spectrum of M82 assuming only one small emitting 
region but with extremely high density of ionized gas. The filling factor of the 
thermal emission may also play an important role in shaping the global spectrum of M82.

\acknowledgements{This work was supported by the Polish Ministry of Science and Higher
Education, grant 3033/B/H03/2008/35.}

\end{document}